# A New Biophysical Metric for Interrogating the Information Content in Human Genome Sequence Variation: Proof of Concept

**James Lindesay [1], Tshela E Mason [2], Luisel Ricks-Santi [3], William Hercules[1], Philip Kurian[1] and Georgia M Dunston [2, 4,]\***

1. Computational Physics Laboratory, Howard University, Washington, DC, 20060, U.S.
   E-mail: jlindesay@howard.edu (J.L.); wmhercules@yahoo.com (W.H.); pjkurian@gmail.com (P.J.K.)
2. National Human Genome Center, Howard University, Washington, DC, 20060, U.S.
   E-mail: tmason@howard.edu (T.E.M.); gdunston@howard.edu (G.M.D.)
3. Cancer Center, Howard University, Washington, DC, 20060, U.S.
   E-mail: lricks-santi@howard.edu (L.R.S.)
4. Department of Microbiology, Howard University, Washington, DC, 20060, U.S.
   E-mail: gdunston@howard.edu (G.M.D.)

\* Author to whom correspondence should be addressed; E-mail: gdunston@howard.edu; Tel: 202-806-7372

**Abstract:** Various studies have shown an association between single nucleotide polymorphisms (SNPs) and common disease. We hypothesize that information encoded in the structure of SNP haploblock variation illumines molecular pathways and cellular mechanisms involved in the regulation of host adaptation to the environment. We developed and utilized the normalized information content (NIC), a novel metric based on SNP haploblock variation. We found that all SNP haploblocks with statistically low information content contained putative transcription factor binding sites and microRNA motifs. We were able to translate a biophysical, mathematical measure of common variants into a deeper understanding of the life sciences through analysis of biochemical patterns associated with SNP haploblock variation. We submit that this new metric, NIC, may be useful in decoding the functional significance of common variation in the human genome and in analyzing the regulation of molecular pathways involved in host adaptation to environmental pathogens.

## 1. Introduction

The human genome is arguably the most sophisticated knowledge system ever discovered, as evidenced by the exquisite information it encodes and communicates via the structure of its DNA sequence. Such information underpins the structure, function, and regulation of complex molecular pathways and network systems transmitted from cell to cell, individual to individual, and generation to generation via the genome. New knowledge derived from sequencing the human genome [1] and researching genome variation [2] challenges traditional views of biological identity and how biology works at the molecular level. The integration of this new knowledge into theoretical models of living systems demands a more complete and comprehensive understanding of the life sciences in general and the science of the human genome in particular. In many respects, the human genome displays features of communication systems based in information theory, such as pattern recognition, data compression, signal processing, and regulation that are also seen from a biophysical perspective of life and living systems.

The concept of information as a basic conserved property of the universe has been successfully demonstrated in the physical sciences from cosmology [3] to telecommunications [4]. In both classical and quantum physics there is a sense that information is conserved. The information content (IC) of an isolated system can be quantified using the fundamental thermodynamic concept of entropy [3]. Complex biophysical systems, like the human genome, are not isolated but rather evolve within external environments which can be assumed to have quasi-static properties. This environmental contact leads to fluctuations of the entropy associated with the system. IC is measured as a difference between the maximum possible entropy and the entropy of a coherently maintained population distribution. In the biological realm, a coherent system maintains its characteristics over generations until perturbed or modified by external forces from the environment.

When applied to whole genome sequence-based biology, we hypothesize that genomic IC can be measured by identifying the "dynamic sites" of the genome and examining variation within and among populations. We identify dynamic sites as single nucleotide polymorphisms (SNPs) for statistical analysis of the genome [5]. Almost all common SNPs have only two alleles and are therefore bi-allelic. SNP haplotype blocks (haploblocks) can be identified, within which the variability of nucleotides can be interrogated and the occurrence of particular combinations can be determined. Analysis of the frequencies of these SNP combinations leads to a statistical distribution of haplotypes within the population. Since the sizes of SNP haploblocks within the genome are of variable lengths, it is nontrivial to directly compare the IC associated with different haploblocks in a meaningful manner. In order to compare the IC among different haploblocks across the genome, as well as among different human populations, a normalized IC (NIC) was developed. The concept for NIC we developed from statistical physics was discovered to be similar to Shannon's concept of *redundancy* [4]. Because our focus is the genome, NIC values here apply to the transmission of information in this biological system. If the NIC value of a SNP haploblock for a population is high compared to 50%, we can deduce that there are likely environmental factors skewing the distribution of haplotype frequencies in the population. Similarly, a low NIC value implies high variability and substantially fewer external factors biasing the haplotype frequency distribution in the population. In particular, a SNP haploblock that is completely homogeneous for a population has identical nucleotides at all dynamic sites for all members, thereby exhibiting no variability in the alleles encoded in that haploblock. Such a *monomorphic* haploblock has a NIC value of unity. Likewise, a population with maximum variation in the alleles will have a NIC value of zero. We assert that populations maintain themselves by establishing coherent SNP haplotype frequencies.

In this paper, we seek to explore the biophysical underpinnings of common variation in the genome. This perspective makes the physics of DNA sequence variation in the human genome relevant in new ways to concepts in biology and biomedical science. In so doing, the intent is to connect the genomic frontiers of biology and the health sciences with the biophysical frontiers of information theory and quantum physics.

## 2. Results and Discussion

*2.1 Distribution of NIC Values for the HLA-DR Gene Region*

Haploview generated 189 haploblocks for the HLA-DR region for an African American population from southwest United States (ASW). As illustrated in Figure 1, the NIC values were distributed between zero and one. There were ten blocks with values proximal to the lower bound (between 0.28 and 0.39), and those blocks were comprised of twenty-three SNPs. Conversely, there were no blocks with NIC values beyond the upper bound (0.86).

**Figure 1.** NIC Values for HLA-DR in an African American Population from the Southwest, US

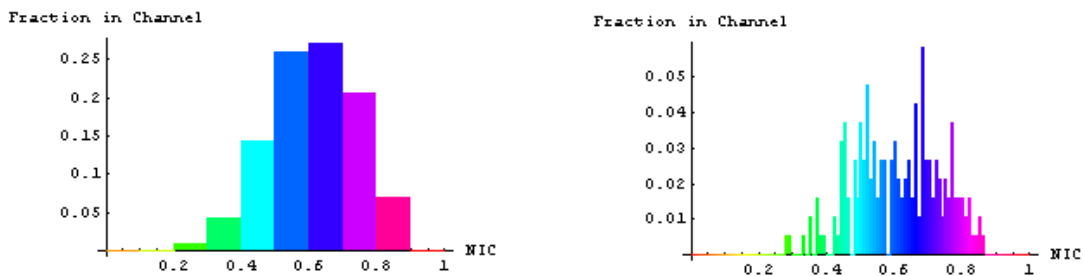

**The height of each bar is the fraction of all NIC values that fall within the channel given by the width of each bar.**

## 2.2 Identification of Putative TFBS and miRNA motifs

Thirty-nine putative transcription factor binding sites (TFBS) were identified using ConSite for the ten blocks proximal to the lower bound which are listed in Table 1.

**Table 1.** Putative TFBS for the Haploblocks proximal to the Lower Bound

| Block ID | ARNT | c-FOS | CHOP | COUP-TF | CREB1 | C-REL | E2F | E4BP4 | FOXA2 | FOXD1 | FOXD3 | FOXF2 | FOXI1 | FOXQ1 | EVI-1 | HEN1 | HLF | IRF-1 | MAX | MEF2 | MYC-MAX | MYF | n-MYC | NF-κB | NRF2 | PAX6 | PBX | P65 | ROR-α1 | ROR-α2 | RUNX1 | RXR-VDR | SAP1 | SOX5 | SOX17 | SPZ1 | STAF | TEF1 | USF |
|---|---|---|---|---|---|---|---|---|---|---|---|---|---|---|---|---|---|---|---|---|---|---|---|---|---|---|---|---|---|---|---|---|---|---|---|---|---|---|---|
| DRB234-174 |  | X |  |  | X |  | X | X | X | X | X | X | X | X |  | X |  |  |  |  |  | X |  |  |  |  |  |  |  |  |  |  | X | X | X |  |  |  | X |
| DRB234-63 | X | X |  |  | X | X | X | X | X | X | X | X | X | X |  | X | X |  | X |  |  | X | X |  |  | X |  |  | X | X | X | X | X | X | X |  |  |  | X |
| DRB234-17 | X | X |  |  | X | X |  |  | X |  |  |  | X |  |  | X |  | X |  |  | X | X | X |  |  | X |  | X |  | X |  | X | X | X | X |  | X | X |  |
| DRB234-182 | X | X |  |  | X | X |  | X | X | X |  | X | X |  |  | X |  |  | X | X |  | X |  |  |  | X |  |  | X |  | X |  | X | X |  |  |  |  | X |
| DRB234-23 |  | X |  | X | X |  |  | X | X | X | X | X | X |  |  | X |  |  |  |  |  | X |  |  | X |  |  |  | X |  | X | X |  |  |  |  |  |  | X |
| DRB234-38 |  | X | X |  |  | X |  | X |  |  |  |  | X |  |  | X |  |  |  | X |  |  |  |  |  |  |  |  | X |  | X | X | X |  |  |  |  |  | X |
| DRB234-132 | X | X |  |  | X | X |  | X | X | X | X | X | X |  |  | X | X |  | X |  |  | X | X |  | X |  |  | X |  | X |  | X | X |  |  |  | X | X |  |
| DRB234-22 |  |  |  |  |  |  |  |  | X | X | X |  | X |  | X |  |  |  |  |  |  |  |  |  |  |  |  |  | X |  |  |  |  | X | X | X |  |  |  |
| DRB234-108 | X | X | X |  |  |  |  | X |  |  | X | X |  | X | X |  | X |  | X | X | X | X | X |  |  |  |  |  |  |  |  | X |  | X | X | X |  |  | X |
| DRB234-40 | X | X |  |  | X |  |  | X | X | X |  | X | X |  |  | X |  |  | X | X |  | X | X | X |  | X | X |  |  |  | X |  |  | X | X | X | X |  | X |

Of the TFBS found, FOXI1, SOX5, and SOX17 sites were present in all ten blocks. Also, there were five SNPs that had TFBS changes when their minor alleles were present (Table 2).

**Table 2.** SNPs proximal to the Lower Bound with TFBS Changes when Minor Allele is Present

| SNPs | Alleles | TFBS Change |
|---|---|---|
| rs9378763 | A>C | Gain of FOXI1 site |
| rs17464136 | C>G | Gain of SOX-17 site |
| rs11970370 | A>C | Loss of SOX-5 site |
| rs6924630 | A>T | Loss of TEF-1 site |
| rs7751939 | C>A | Loss of IRF-1 and gain of FOXA2, FOXD3 & FOXI1 sites |

When the blocks proximal to the lower bound were scanned for miRNA motifs using miRBase, we found that all ten blocks had miRNA motifs present (Table 3).

**Table 3.** Characterized miRNAs Located proximal to the Lower Bound

| Block ID | miRNA |
|---|---|
| DRB234-23 | let-7e |
|  |  |
| DRB234-17 | miR-16 |
|  | miR-93 |
|  | miR-222 |
|  |  |
| DRB234-63 | miR-29 |
|  |  |
| DRB234-132 | miR-142-p |
|  | miR-548 |
|  |  |
| DRB234-38 | miR-21 |

*2.3 Location of SNPs proximal to the Lower Bound*

All twenty-three SNPs proximal to the lower bound were located in intergenic regions (Table 4).

**Table 4.** Location of SNPs proximal to the Lower Bound

| Block ID | NIC Value | SNPs | SNP Location |
|---|---|---|---|
| DRB234-174 | 0.28 | rs9378385 | Intergenic Region |
| | | rs9503746 | Intergenic Region |
| DRB234-63 | 0.29 | rs1028380 | Intergenic Region |
| | | rs7774941 | Intergenic Region |
| DRB234-17 | 0.33 | rs1890366 | Intergenic Region |
| | | rs2788212 | Intergenic Region |
| DRB234-182 | 0.35 | rs9405676 | Intergenic Region |
| | | rs9378389 | Intergenic Region |
| DRB234-23 | 0.35 | rs7751939 | Intergenic Region |
| | | rs6597267 | Intergenic Region |
| | | rs2317217 | Intergenic Region |
| DRB234-38 | 0.37 | rs11970370 | Intergenic Region |
| | | rs845896 | Intergenic Region |
| DRB234-132 | 0.37 | rs6924630 | Intergenic Region |
| | | rs9378763 | Intergenic Region |
| | | rs6596945 | Intergenic Region |
| DBR234-22 | 0.37 | rs9392155 | Intergenic Region |
| | | rs9505192 | Intergenic Region |
| | | rs9505153 | Intergenic Region |
| DRB234-108 | 0.38 | rs2449447 | Intergenic Region |
| | | rs1773015 | Intergenic Region |
| DRB234-40 | 0.39 | rs1764136 | Intergenic Region |
| | | rs845883 | Intergenic Region |

*2.4 The Significance of the Information Content in the Natural Variation of the HLA-DR Region*

The advent of geographically-defined, population-based, genome-wide variation resources such as the haplotype map (i.e. HapMap) has opened a new era in human population genetics. Single nucleotide polymorphisms (SNPs), the most common type of natural variation in the human genome, offer an unprecedented opportunity to investigate evolutionary forces that have shaped human genome variation in natural populations. Information content (IC), as a new metric grounded in the biophysical matrix of DNA-sequence based biology, has been used to explore the biomedical significance of natural variation in the human genome. As the most polymorphically expressed biological system, the HLA region was examined for proof of concept that natural variation encodes fundamental information about the biology of host adaptive mechanisms in response to environmental pathogens. The data on the normalized information content (NIC) of SNP haploblocks in the HLA-DR region relate natural variation to pathways of innate immunity. The molecules and cells of the innate immune system are the first responders to environmental disturbances of homeostasis. This system can initiate the inflammatory response by activating the cellular release of cytokines, chemokines, reactive oxygen (ROS), and reactive nitrogen (RNS) species. However, when the inflammatory response is sustained at a chronic level, these molecules can inflict a variety of damaging effects.

*2.5 The Transcription Factor Binding Sites Present in Ten Blocks Proximal to the Lower Bound*

In our analysis, the transcription factor binding sites (TFBS) present in all ten of the blocks proximal to the lower bound have been reported to be regulated by the p38 *mitogen activated protein kinase* (MAPK) and Wnt (wingless*)* pathways[6,7,8]. These two pathways are activated in response to oxidative stress and inflammation [9, 10]. p38 is a kinase that is localized to the cytoplasm until activated, when it translocates to the nucleus. Its activity is critical for normal immune and inflammatory responses. p38 is activated by macrophages, neutrophils, and T cells in response to cytokines, chemokines, and bacterial lipopolysaccharide (LPS) [11]. It is known to phosphorylate its cellular targets such as the following transcription factors: ATF-1 and -2, MEF2A, Sap-1, Elk-1, NF-κB, Ets-1, and p53 [11]. An interesting feature of the p38 pathway is that it can regulate the Wnt pathway when activated by ROS and other stressors. Wnt signaling controls a variety of signal transduction pathways that involve protein kinases, caspases, NF-κB, GSK-3β, iNOS and FOXes [9, 12, 13].

Thus, it is possible that cross-talk between the p38 and Wnt pathways represents a network that has formed in response to oxidative stress. This proposed network would be advantageous to a population under constant challenge in a tropical environment with a plethora of pathogenic agents, such as *Schistosoma mansoni* (*S. mansoni)*. In innate immunity, dendritic cells (DCs) exposed to helminth products (such as Lacto-N-Fucopentaose III (LNFPIII), a milk sugar containing the Lewis$^x$ trisaccharide found in the schistosome egg antigen (SEA)), have been reported to activate NF-κB by stimulating its nuclear translocation[14]. It is worth noting that Lewis structures can occur on both N-glycan and mucin-type O-glycan cores, and these fucosylated glycans have been involved in many functions, like selectin recognition [15]. Selectins are known for mediating extravasation of leukocytes and lymphocytes, pathogen adhesion, and modulation of signal transduction pathways [16]. The hallmark of *S. mansoni* infection is the switching of the host immune response from Th1 to Th2, resulting in the persistent survival of the parasite. One of the key components involved in modulating the host immune response from Th1 to Th2 is NF-κB. Studies conducted by Goodridge *et al*. and Thomas *et al*. [17, 18], found that neither SEA nor LNFPIII-dextran pulsed NF-κB $^{-/-}$ DCs were able to induce a Th2 response. Interestingly, increased levels of IL-4, a Th2 cytokine, have been demonstrated in murine schistosomiasis to control the generation of reactive oxygen and nitrogen intermediates in the liver [19].

*2.6 Putative miRNA Structures Present in Ten Blocks Proximal to the Lower Bound*

Additionally, our analysis showed that all the blocks proximal to the lower bound contained putative miRNA structures. This is consistent with findings of miRNAs in intronic and intergenic regions [20]. There were four blocks that contained well characterized miRNAs (Table 3). These miRNAs have been identified as tumor suppressors in a multitude of cancers [21, 22]. Furthermore, miRNA expression has been shown to regulate the inflammatory response. For example, O'Connell *et al*. [23] reported that an increase in miR-155 and a decrease in let-7e levels enhanced the response of Akt$^{-/-}$ macrophages to LPS.

*2.7 The SNP Variation in the Region Proximal to the Lower Bound*

Particularly noteworthy was the SNP variation proximal to the lower bound, which reflected that five SNPs had TFBS changes when their minor alleles were present (Table 4). rs9378763 and rs1764136 had a gain of FOXI1 and SOX-17 sites, respectively. rs11970970 and rs6924630 had a loss of SOX-5 and TEF-1 sites, respectively. rs7751939 was the only SNP whose minor allele resulted in loss of an IRF-1 site and gain of FOXA2, FOXD3 and FOXI1 sites. These FOX transcription factors have been shown to activate and be activated by the Wnt pathway [24, 8]. In the study by Zhang *et al*. [25], SOX-17 was shown to negatively regulate the Wnt pathway via suppression of β-catenin/T-cell factor-regulated transcription. IRF-1 is constitutively expressed in various cell types and induces the expression of pro-inflammatory cytokines in response to the activation of pathogen recognition receptors, such as the Toll-like receptor and nucleotide-binding oligomerization domain (NOD)-like receptor families [26]. Also, IRF1 transcriptionally targets a number of genes, and is required for Th1 differentiation of interferon (IFN)-stimulated macrophages. When IRF-1 is absent, the induction of Th2-type immune responses occurs [27]. In addition, when p38 is activated by IFN, it contributes to the phosphorylation of NF-κB, AP-1, IRF-1, IRF-4, IRF-8, and PU.1 [28]. It has been reported that transcription enhancer factor 1 (TEF-1) directly binds to poly (ADP-ribose) polymerase 1(PARP-1) which is known to participate in DNA repair processes [29]. Under conditions of oxidative stress, the activation of PARP1 results in greater expression of AP-1 and NF-κB-dependent genes [30]. Also, in a study conducted by Braam *et al.*[31] using endothelial cells, SOX-5 had more pronounced representation in genes regulated by nitrous oxide (NO) than the other transcription factors studied. Interestingly, in *falciparum* malaria NO inhibits the adhesion of parasitized red cells to vascular endothelium [32]. Hence, it is possible that not only has schistosomiasis acted as an environmental stressor in shaping the allelic variation in the proposed p38-Wnt compensatory network, but so has malaria. It is intriguing that independent of the potential gains and losses of transcription factor binding sites, there is continued regulation of the oxidative stress process irrespective of specific allele selection.

*2.8 Comparison of NIC Values Proximal to the Lower Bound to Those Values Most Proximal to the Upper Bound*

We also assessed the performance of the measure by comparing blocks with NIC values most proximal to the upper bound with those most proximal to the lower bound. It is noteworthy that more than 96% of SNPs in the blocks proximal to the upper bound were located in genic regions, in contrast to the SNPs in the blocks proximal to the lower bound, *none* of which were found in genic regions. The identification of genic haploblocks with high information content and a more in-depth assessment of the biological significance of the entire NIC distribution are being investigated.

**3. Experimental Section**

*3.1 Derivation of the Normalized Information Content Equation*
The degree of variability within a SNP haploblock population can provide a measure of the maintained order associated with that haploblock. SNP haplotype diversity will vary across different SNP haploblocks. Each population group is defined by the maintained order of its SNP haplotype diversity within the SNP haploblock structure; however the latter might be defined. Thus, haplotype diversity is herein reflected in the frequencies with which the SNP haplotypes occur within a given haploblock structure.

In order to provide a meaningful comparison of the information content among different regions of the genome as well as amongst different populations, the normalized information content (NIC) parameter was developed. NIC measures the difference between the entropy and the maximum possible entropy of a SNP haploblock within a given population. Since we expect that the external environment will significantly influence the state of the genome, we choose a form for the entropy measure as illustrated in equation 1.

$$S_{A,coherent} = -k \sum_{j}^{N_A} P^A_j \log_2 P^A_j . \qquad (Eq.\ 1)$$

where $P^A_j$ represents the probability or frequency with which a particular SNP haplotype j occurs within the particular haploblock A, and $N_A = 2^{N^A_{SNPs}}$ represents the number of mathematically possible

SNP combinations for $N^A_{SNPs}$ active biallelic SNP sites. For our purposes, $S_{coherent}$ has the potential of being an additive genostatic parameter that can be used to quantify the information in a system. Since all probabilities are non-negative, the minimum value this entropy can take occurs when one of the probabilities itself is unity. This defines a *homogeneous* population yielding $S_{A,min}=0$. The maximum value this entropy can take occurs when all the probabilities are equal, $P^A_j = \frac{1}{N_A}$, defining an informationally *gray population*. In this case, one obtains the result for the entropy as $S_{A,max} = k \log_2 N_A = k N^A_{SNPs}$. This represents the mathematical maximum of entropy for the SNP haploblock A across all human populations, giving a universal upper bound for this value. A given population will generally not have all possible SNP combinations expressed as viable SNP haplotypes, so that a population made up of equal frequencies for just the expressed haplotypes would not represent a universal maximum across populations. However, some of the SNP combinations that are not expressed in the populations *do* contribute to the information content of the haploblock. For these reasons, the chosen form for the maximum entropy is both universal and complete. Since the maximum entropy represents the upper limit that any entropy can attain, genomic information can be expected to relate to the difference between the maximum entropy of a gray population and the coherent frequency distribution maintained by a given population. However, the number of SNPs haplotypes that completely describe all populations varies for different regions of the genome. The information measure is most useful when it represents a dimensionless parameter that can be used to compare the information content of different SNP haploblocks as well as the same block amongst different populations. We have therefore chosen to normalize our measure of information content in a manner that gives zero for a gray population with no information content and unity for a homogeneous population that has a single maintained SNP haplotype. This can be mathematically expressed for haploblock A as follows:

$$NIC_A = \frac{S_{A,max} - S_{A,coherent}}{S_{A,max}} = \frac{N^A_{SNPs} + \sum_{j=1}^{2^{N^A_{SNPs}}} P^A_j \log_2 P^A_j}{N^A_{SNPs}}. \quad \text{(Eq. 2)}$$

This information measure is bounded by $0 \leq NIC \leq 1$, allowing for an informational comparison of apples to oranges. Normalization of the information metric therefore provides a means to interrogate genomic information from different regions of the genome. Besides limiting the range of the *NIC*, the maximum entropy state of specified block A is common to all populations; only the frequency distribution varies amongst the populations. This dimensionless form allows one to gain considerable insights into genomic information without a need for detailed information about the dynamics and history of the genome or knowledge of the form of all genomic parameters. The NIC is similar to an independently derived from [33] where an entropic measure was used to construct haploblocks rather than interrogate the information content of SNP haploblock variation.

*3.2 Analysis of NIC for the Human Leukocyte Antigen-Disease Related (HLA-DR) region of the Major Histocompatibility Complex (MHC)*

The human MHC encoding the HLA system is the most highly expressed polymorphic system in the genome, and it plays an essential role in regulation of the immune response in host adaptation to environmental stimuli. It is a strong genomic marker of historical changes in environmental conditions. The NIC values were calculated for the HLA-DR region located on chromosome 6 between positions 415,611 and 3,908,995 (HLA-DRA1, HLA-DRB1, HLA-DRB5) and between positions 32,515,990 and 32,663,637 (HLA-DRB2, HLA-DRB3, HLA-DRB4). SNP haploblocks were constructed using the confidence interval algorithm [34] in Haploview v 4.2 from HapMap phase III data on the ASW population (N=98). Haploview uses a two marker expectation-maximization algorithm with a partition-ligation approach which creates highly accurate population frequency estimates of the phased haplotypes based on the maximum-likelihood as determined from the unphased input [35, 36]. The confidence interval algorithm defines the haploblock as a region over which a very small proportion (<5%) of comparisons among "informative SNP pairs" shows strong evidence of historical recombination and within which independent measures of pairwise linkage disequilibrium (LD) did not decline substantially with distance. Informative SNP pairs were in "strong LD" if the one-sided upper 95% confidence bound on the pairwise correlation factor D' was > 0.98 and the lower bound above 0.7. Conversely, "strong evidence of historical recombination" was defined by an upper confidence bound on D' less than 0.9 [34]. The NIC values for the various blocks were collectively plotted in order to assess the statistical features of the distribution. We defined outliers on the distribution by identifying those values proximal to or beyond two root mean squared (RMS) deviations ($2\sigma$) of the mean, 0.61 (RMS ($\sigma$) =0.13; standard error=0.01). Root mean squared deviation is defined by $\sigma = \sqrt{<(p-<p>)^2>}$, where $<p>$ represents the mean of the distribution with discrete elements $p_j$. Lastly, all of these regions were scanned for potential regulatory elements using the publicly available bioinformatics tools ConSite and miRBase.

## 4. Conclusion

This paper has introduced a biophysical metric for analyzing the information content of SNP haploblock variation. NIC values in the HLA-DR region highlighted common variants involved in regulation of host immunity to environmental stressors. This supports our hypothesis that information encoded in the structure of SNP haploblock variation can elucidate molecular pathways and cellular mechanisms involved in the regulation of host adaptation to the environment. Using our analysis, p38 and Wnt, are proposed as a communication network connected by transcription factors and miRNAs in population adaptation to pathogens. Since the genetic variation highlighted by NIC values is in a representative sample of the population, its relevance in disease association studies remains to be determined. Because our motivation has been to use common variation in a reference population to interrogate the biology of health, the further understanding of disease using this approach would be a by-product. Finally, NIC values derived from common variation in the HLA-DR region suggest its involvement with regulation of innate immune mechanisms.

**Acknowledgments**

The research is supported in part by NIH Grants NCRR 2 G12 RR003048 from the RCMI Program, Division of Research Infrastructure; NIGMS S06 GM08016, and NCI 5U54 CA091431.